\RequirePackage{ifpdf}
\ifpdf 
\documentclass[pdftex]{sigma}
\else
\documentclass{sigma}
\fi

\newcommand{\simr}{\hspace{.3em}\raisebox{.4ex}{$>$}\hspace{-.75em}
 \raisebox{-.7ex}{$\sim$}\hspace{.3em}}
 \newcommand{\siml}{\hspace{.3em}\raisebox{.4ex}{$<$}\hspace{-.75em}
 \raisebox{-.7ex}{$\sim$}\hspace{.3em}}

\begin{document}

\allowdisplaybreaks

\renewcommand{\PaperNumber}{018}

\FirstPageHeading

\renewcommand{\thefootnote}{$\star$}

\ShortArticleName{Lattice Field Theory with  the Sign Problem and the Maximum Entropy Method}

\ArticleName{Lattice Field Theory with  the Sign Problem \\ and the Maximum Entropy Method\footnote{This paper is a
contribution to the Proceedings of the O'Raifeartaigh Symposium on
Non-Perturbative and Symmetry Methods in Field Theory (June
22--24, 2006, Budapest, Hungary). The full collection is available
at \href{http://www.emis.de/journals/SIGMA/LOR2006.html}{http://www.emis.de/journals/SIGMA/LOR2006.html}}}


\Author{Masahiro IMACHI~$^\dag$, Yasuhiko SHINNO~$^\ddag$ and Hiroshi  YONEYAMA~$^{\S}$}
\AuthorNameForHeading{M. Imachi, Y. Shinno and H. Yoneyama}
\Address{$^\dag$~Kashiidai, Higashi-ku, Fukuoka, 813-0014, Japan}

\Address{$^\ddag$~Takamatsu National College of Technology, Takamatsu 761-8058,  Japan}

\Address{$^{\S}$~Department of Physics, Saga University, Saga, 840-8502, Japan}
\EmailD{\href{mailto:yoneyama@cc.saga-u.ac.jp}{yoneyama@cc.saga-u.ac.jp}}

\ArticleDates{Received September 30, 2006, in f\/inal form January
19, 2007; Published online February 05, 2007}

\Abstract{Although  numerical simulation in lattice f\/ield theory is one of the most ef\/fective tools to study
  non-perturbative properties of f\/ield theories, it faces serious obstacles coming  from  the  sign problem in some theories such as
  f\/inite density QCD and lattice f\/ield theory with  the
  $\theta$ term.  We  reconsider this problem  from the point of view of the maximum
  entropy method.}


\Classification{65C05; 65C50; 68U20; 81T25; 81T80}

\section{Introduction}

\looseness=-1
   Lattice f\/ield theory is a powerful method to study non-perturbative aspects of  quantum f\/ield theo\-ry.
 Although  numerical simulation is one of the most ef\/fective tools to study
  non-perturba\-ti\-ve properties of f\/ield theories, it  faces serious obstacles in theories such as
  f\/inite density QCD and lattice f\/ield theory with  the
  $\theta$ term.  This is because  the Boltzmann weights are  complex and this makes it
  dif\/f\/icult to perform  Monte Carlo (MC) simulations on a
  Euclidean lattice. This is the complex action problem, or the sign
  problem. In the present
talk, we review   the analysis of the sign problem based on    the maximum
  entropy method (MEM)~\cite{rf:Bryan,rf:JG,rf:AHN}.
  For details, refer to~\cite{rf:ISY, rf:ISY2, rf:ISY3}.  The
   MEM is well known as a powerful tool for
  so-called  ill-posed problems, where the number of parameters to be
  determined is much larger than the number of data points. It has
  been  applied to
  a wide range of f\/ields,  such as radio astrophysics and
  condensed matter physics.

   In this talk  we deal only with lattice f\/ield theory with  the
  $\theta$ term.     It is believed  that the $\theta$ term could af\/fect the
  dynamics at low energy and the vacuum structure of QCD, but it is known
  from experimental evidence that the value of $\theta$ is strongly
  suppressed in Nature. From the theoretical point of view, the reason
  for this is not clear yet. Hence, it is important to study the properties
  of QCD with the
  $\theta$ term to clarify the structure of the QCD
  vacuum~\cite{rf:tHooft, rf:CR,rf:CR1}.
  For  theories with the $\theta$ term,
  it has been  pointed out that  rich phase
  structures could be realized in
  $\theta$ space. For example, the phase structure
  of the  $Z(N)$ gauge model was investigated using free energy arguments,  and
  it was found that
  oblique conf\/inement phases could occur~\cite{rf:CR,rf:CR1}.
  In CP$^{N-1}$ models, which have
  several dynamical properties in common with QCD, it has been  shown that a
  f\/irst-order phase transition exists at
   $\theta=\pi$~\cite{rf:Seiberg,rf:BRSW, rf:Wiese}.

 In order to circumvent the sign  problem, the following method is
  conventionally employed~\cite{rf:BRSW, rf:Wiese}.   The partition function
${\cal Z}(\theta)$ can
  be obtained by Fourier-transforming the topological charge distribution
  $P(Q)$, which is calculated with a  real positive Boltzmann weight:
\begin{gather*}
  {\cal Z}(\theta)=\frac{\int[d\bar{z}dz]e^{-S+i\theta\hat{Q}(\bar{z},z)}}
   {\int[d\bar{z}dz]e^{-S}}\equiv\sum_{Q}e^{i\theta Q}P(Q),
\end{gather*}
  where
\begin{gather}
  P(Q)\equiv \frac{\int[d\bar{z}dz]_Qe^{-S}}{\int d\bar{z}dz e^{-S}}.
   \label{eqn:Pq}
\end{gather}
  The measure $[d\bar{z}dz]_Q$ in equation~(\ref{eqn:Pq}) is such  that the
integral is
  restricted to conf\/igurations of the f\/ield $z$ with  topological
charge $Q$. Also,
  $S$ represents the  action.

  In the study  of CP$^{N-1}$ models, it is known that this algorithm
  works well for a small lattice volume $V$ and in the strong coupling
  region~\cite{rf:Seiberg, rf:BRSW,rf:PS, rf:IKY}.
  As the volume increases or in the weak coupling region,  however,
  this strategy too suf\/fers from the
  sign problem for $\theta\simeq\pi$. The error in
  $P(Q)$ masks the true values of ${\cal Z}(\theta)$ in the vicinity of
  $\theta=\pi$, and this results in  a f\/ictitious signal of a phase
  transition~\cite{rf:PS, rf:IKY}. \  This is called `f\/lattening', because the free
  energy becomes almost f\/lat for $\theta$ larger than a certain value.
  This problem  could be remedied by reducing the error in  $P(Q)$.
  This, however, is hopeless,  because  the
  amount of data needed to reduce the error to a given level
  increases exponentially with $V$.

   Here, we are interested in whether the MEM can be
   applied ef\/fectively to
   the study of the $\theta$ term and  reconsider
  the f\/lattening phenomenon of the free energy in terms of the MEM.
  The MEM is based upon  Bayes' theorem. It derives the most probable
  parameters by utilizing data sets and our knowledge about these
  parameters in terms of the probability. The probability distribution,
  which is called the posterior probability, is given by the product of the
  likelihood function and the prior probability. The latter is
  represented by the Shannon-Jaynes entropy, which plays an important
  role to guarantee the uniqueness of the solution, and the former is
  given by $\chi^2$. It should be noted that artif\/icial
  assumptions are not needed in  the calculations, because
  the determination of a unique solution is carried out  according to
  probability theory.
   Our task is to determine  the image for which  the
  posterior probability is maximized.

  We present the results for the analysis by
i) using mock data and ii) using the MC data.
For the former, we use the Gaussian form of $P(Q)$.  The Gaussian form
  is  realized in many  cases,   such as  the strong coupling region
   of the CP$^{N-1}$ model and the
  2-d $U(1)$ gauge model.    For using MC data,   we simulate  the CP$^{N-1}$ model
  and  apply the MEM to the obtained data.
    This paper is organized as follows. In the following section, we
  give an
  overview of the origin of f\/lattening and   summarize the procedure for the analysis of the MEM.
  The results obtained
by use of
  the MEM are presented in Section~\ref{sec:result}.

 \section{Flattening and MEM}
 \subsection{Flattening}
   The free energy density $f(\theta)$ is calculated by
 Fourier-transforming $P(Q)$ obtained by MC simu\-lation.  Let us call this method the FTM.
 The quantity $f(\theta)$ is def\/ined as
\begin{gather*}
 f(\theta)=-\frac{1}{V}\ln\sum_QP(Q)e^{i\theta Q}, 
\end{gather*}
 where $V=L^2$, the square of the lattice size.

 The MC data for $P(Q)$ consist of the true value, ${\tilde P}(Q)$,
 and its error, $\Delta P(Q)$. When the error  at $Q=0$
  dominates because of the exponential damping of $P(Q)$,
 $f(\theta)$ is closely approxima\-ted~by
\begin{gather*}
 f(\theta)\simeq -\frac{1}{V}\ln\big[e^{-V{\tilde f}(\theta)}+\Delta P(0)
                 \big],
\end{gather*}
 where ${\tilde f}(\theta)$ is the true value of $f(\theta)$.
 Because  ${\tilde f}(\theta)$ is an increasing function of $\theta$,
 $\Delta P(0)$ dominates for large values of $\theta$. If $|\Delta
 P(0)|\simeq e^{-V{\tilde f}(\theta)}$ at $\theta=\theta_{\rm f}$, then
 $f(\theta)$ becomes almost f\/lat for $\theta\simr \theta_{\rm f}$. This
 is called ``f\/lattening of the free energy density", and  it has been  misleadingly
 identif\/ied as a f\/irst order phase transition,
 because the f\/irst derivative of $f(\theta)$  appears to jump  at
 $\theta=\theta_{\rm f}$.
 To avoid  this problem,
 we must carry out  the order of  $e^V$ measurements in the FTM.

  \subsection{MEM}
 In this subsection, we brief\/ly explain the MEM in terms of the $\theta$
 term.
 In a parameter inference,  such as the $\chi^2$ f\/itting, the inverse
 Fourier transform
\begin{gather}
 P(Q)=\int_{-\pi}^{\pi}\frac{d\theta}{2\pi}{\cal Z}(\theta)e^{-i\theta Q}
  \label{eqn:invfourier}
\end{gather}
 is used. In the  numerical calculations, we use the discretized version
 of equation~(\ref{eqn:invfourier}); $P(Q)=\sum\limits_n K_{Q,n}{\cal Z}_n$,
 where $K_{Q,n}$ is the Fourier integral kernel and ${\cal Z}_n\equiv
 {\cal Z}(\theta_n)$. In order for  the continuous function ${\cal Z}(\theta)$
  to be reconstructed, a suf\/f\/icient number of values  of
 $\theta$, which we denote  by~$N_{\theta}$, is required so that the relation $N_{\theta}>N_Q$ holds,
  where $N_Q$ represents the number of data points in
 $P(Q)$ ($Q=0$, $1,\dots,N_Q-1$). A straightforward application of the
 $\chi^2$ f\/itting to the case $N_{\theta}>N_Q$ leads to degenerate
 solutions.
 This is an ill-posed problem.

 The maximum entropy method is suitable  to solve  this
 ill-posed problem, yielding  a unique solution.
 The MEM is based upon Bayes' theorem, expressed as
\begin{gather*}
 {\rm prob}({\cal Z}(\theta)|P(Q),I)=
  \frac{{\rm prob}(P(Q)|{\cal Z}(\theta),I)\,{\rm prob}({\cal Z}(\theta)|I)}
  {{\rm prob}(P(Q)|I)}, 
\end{gather*}
 where ${\rm prob}(A|B)$ is the conditional probability
 that $A$ occurs under the condition that $B$ occurs.
 The posterior probability ${\rm prob}({\cal Z}(\theta)|P(Q),I)$ is the
 probability that the partition function ${\cal Z}(\theta)$ is realized  when the
 MC data $\{P(Q)\}$ and prior information $I$ are given.
 The likelihood function ${\rm prob}(P(Q)|{\cal Z}(\theta),I)$ is given
 by
\begin{gather*}
 {\rm prob}(P(Q)|{\cal Z}(\theta),I)=\frac{1}{X_L}e^{-\frac{1}{2}\chi^2},
\end{gather*}
 where $X_L$ is a normalization constant and $\chi^2$ is a standard
 $\chi^2$ function. \par
 The probability ${\rm prob}({\cal Z}(\theta)|I)$ is given in terms of  an
 entropy $S$ as
\begin{gather*}
 {\rm prob}({\cal Z}(\theta)|I)
=\frac{1}{X_s(\alpha)}e^{-\alpha S}, 
\end{gather*}
 where $\alpha$ and $X_S(\alpha)$ are a positive parameter and an
 $\alpha$-dependent normalization constant, respectively. As $S$, the
 Shannon--Jaynes entropy is conventionally employed:
\begin{gather*}
 S=\sum_{n=1}^{N_{\theta}}\left[{\cal Z}_n-m_n
    -{\cal Z}_n\ln\frac{{\cal Z}_n}{m_n}\right].
\end{gather*}
Here  $m_n\equiv m(\theta_n)$ represents  a default model.

 The posterior probability ${\rm prob}({\cal Z}_n|P(Q),I)$, thus, is given by
\begin{gather*}
 {\rm prob}({\cal Z}_n|P(Q),I,\alpha,m)=\frac{1}{X_LX_s(\alpha)}
  e^{-\frac{1}{2}\chi^2+\alpha S}\equiv\frac{e^{W[{\cal Z}]}}{X_LX_s(\alpha)}.
\end{gather*}
 For the prior information $I$, we impose the criterion
\begin{gather*}
 {\cal Z}_n>0,  
\qquad \mbox{so that \ \ ${\rm prob}({\cal Z}_n\leq 0\,|\,I,\alpha,m)=0$}.
 \end{gather*}
  The most probable image of ${\cal Z}_n$, denoted as ${\hat {\cal
  Z}}_n$, is calculated according to the following
  procedures~\cite{rf:AHN,rf:ISY}.
\begin{enumerate}\itemsep=0pt
 \item Maximizing $W[{\cal Z}]$ to obtain the most probable image
       ${\cal Z}_n^{(\alpha)}$ for a given $\alpha$:
\begin{gather*}
 \frac{\delta}{\delta{\cal Z}_n}W\left[{\cal Z}\right]\Big|_
  {{\cal Z}={\cal Z}^{(\alpha)}}=
  \frac{\delta}{\delta{\cal Z}_n}\left(-\frac{1}{2}\chi^2+\alpha S
  \right)\Big|_{{\cal Z}={\cal Z}^{(\alpha)}}=0. 
\end{gather*}
 \item Averaging ${\cal Z}^{(\alpha)}_n$ to obtain the
       $\alpha$-independent most probable image ${\cal Z}_n$:
\begin{gather*}
 {\hat{\cal Z}}_n=\int d\alpha~{\cal Z}^{(\alpha)}_n~
  {\rm prob}(\alpha|P(Q),I,m).
\end{gather*}
       The range of integration is determined so that  the relation
\[
{\rm prob}(\alpha|P(Q),I,m)\geq{\rm prob}
       ({\hat \alpha}|P(Q),I,m)/10
\]
       holds, where
       ${\rm prob}(\alpha|P(Q),I,m)$
       is maximized at $\alpha={\hat \alpha}$.
 \item Error estimation:
       The error of the most probable output image ${\hat {\cal Z}}_n$ is
       calculated as the uncertainty of the image, which takes into
       account the correlations of the images ${\hat {\cal Z}}_n$ among
       various  values of $\theta_n$:
\begin{gather}
 \langle(\delta{\hat{\cal Z}}_n)^2\rangle\equiv\int d\alpha~
  \langle(\delta{\cal Z}_n^{(\alpha)})^2\rangle~{\rm prob}(\alpha|P(Q),I,m).
 \label{eqn:uncertainty}
\end{gather}
 Here  $\delta{\hat{\cal Z}_n}$ and $\delta{\cal Z}_n^{(\alpha)}$
       represent the error in ${\hat{\cal Z}}_n$ and that in
       ${\cal Z}_n^{(\alpha)}$, respectively.
\end{enumerate}

\section{Results}
\label{sec:result}
\subsection{Mock data: Gaussian}
   Firstly, we present the results of the analysis by using mock data.
   For this,  we use  the Gaus\-sian~$P(Q)$ as
\[
   P(Q)=A \exp\left[-\frac{c}{V}Q^2\right], 
\]
   where, in the case of the 2-d $U(1)$ gauge model, $c$ is a constant depending
   on the inverse coupling constant $\beta$,  and $V$ is the lattice
   volume.
   The  constant  $A$ is  f\/ixed  so that   $\sum\limits_Q P(Q)=1$.
   The distribution $P(Q)$ is analytically transformed by use of the
Poisson sum
   formula into the partition function
\begin{gather}
   {\cal Z}_{\rm pois}(\theta)=A\sqrt{\frac{\pi V}{c}}\sum_{n=-\infty}^{\infty}
    \exp\biggl[-\frac{V}{4c}(\theta-2\pi n)^2\biggr]. \label{eqn:poissonsum}
\end{gather}
   To  prepare  the mock data, we add noise with  variance
   $\delta \times P(Q)$ to the Gaussian  $P(Q)$. In  the analysis, we
   consider sets of
   data  with various values of  $\delta$ and study the ef\/fects of
$\delta$.

\begin{figure}[h]
\centerline{\includegraphics[width=13cm]{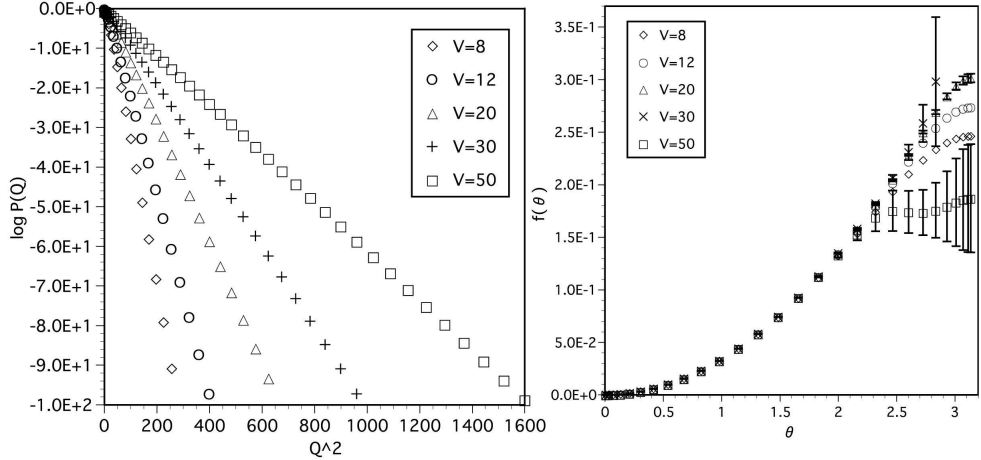}}
\vspace{-3mm}
\caption{Gaussian topological charge distribution
and corresponding $f(\theta)$  obtained by using the Fourier method (FTM) for  various
lattice volumes~\cite{rf:ISY}. The parameter $\delta$ is chosen to be 1/400.}
\label{fig:Pq}\vspace{-1mm}
\end{figure}

In  Fig.~\ref{fig:Pq}  the Gaussian topological charge distribution  and corresponding $f(\theta)$  obtained by using the FTM  are shown  for  various
lattice volumes.  For small volumes, the behavior of  $f(\theta)$ is smooth.
For large volume ($V=50$), however, clear f\/lattening is  observed.
For $V=30$, some data are missing. This is due to the fact that  $ {\cal Z}(\theta)$  could take negative values because of large errors. This is also called f\/lattening, because its origin is the same as  that stated above.

\begin{figure}[h]
\centerline{\includegraphics[width=7.0cm]{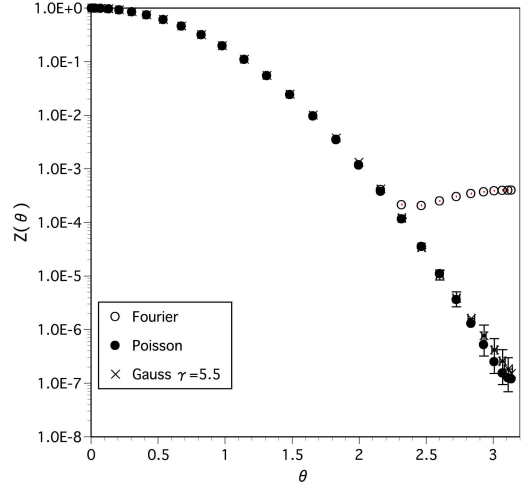}}
\vspace{-3mm}
\caption{$\hat {{\cal Z}}(\theta)$ (crosses) with the error bars for  the
Gaussian default model with $\gamma=5.5$~\cite{rf:ISY}. Here, $V=50$. Compared to
the  result of  the FTM (circles), a remarkable improvement is clearly seen.\label{fig:Z_quad}}\vspace{-2mm}
\end{figure}

For the $P(Q)$   data corresponding to   small volumes without f\/lattening, the MEM successfully reproduces $f(\theta$).
Fig.~\ref{fig:Z_quad} displays $ {\cal Z}(\theta)$ for $V=50$ by using the MEM.
Here, the Gaussian default  $m(\theta)=\exp\left[-\gamma\frac{\ln
 10}{\pi^2}\theta^2\right]$ with $\gamma=5.5$ is used.
The result  of the MEM does not show f\/lattening but smooth behavior, in agreement with  the exact result calculated by equation~(\ref{eqn:poissonsum}) (Poisson).      In contrast, the FTM yields f\/lattening.
See \cite{rf:ISY} for details.

\subsection{MC data}
\begin{figure}[t]
 \begin{minipage}{75mm}
   \includegraphics[width=75mm]{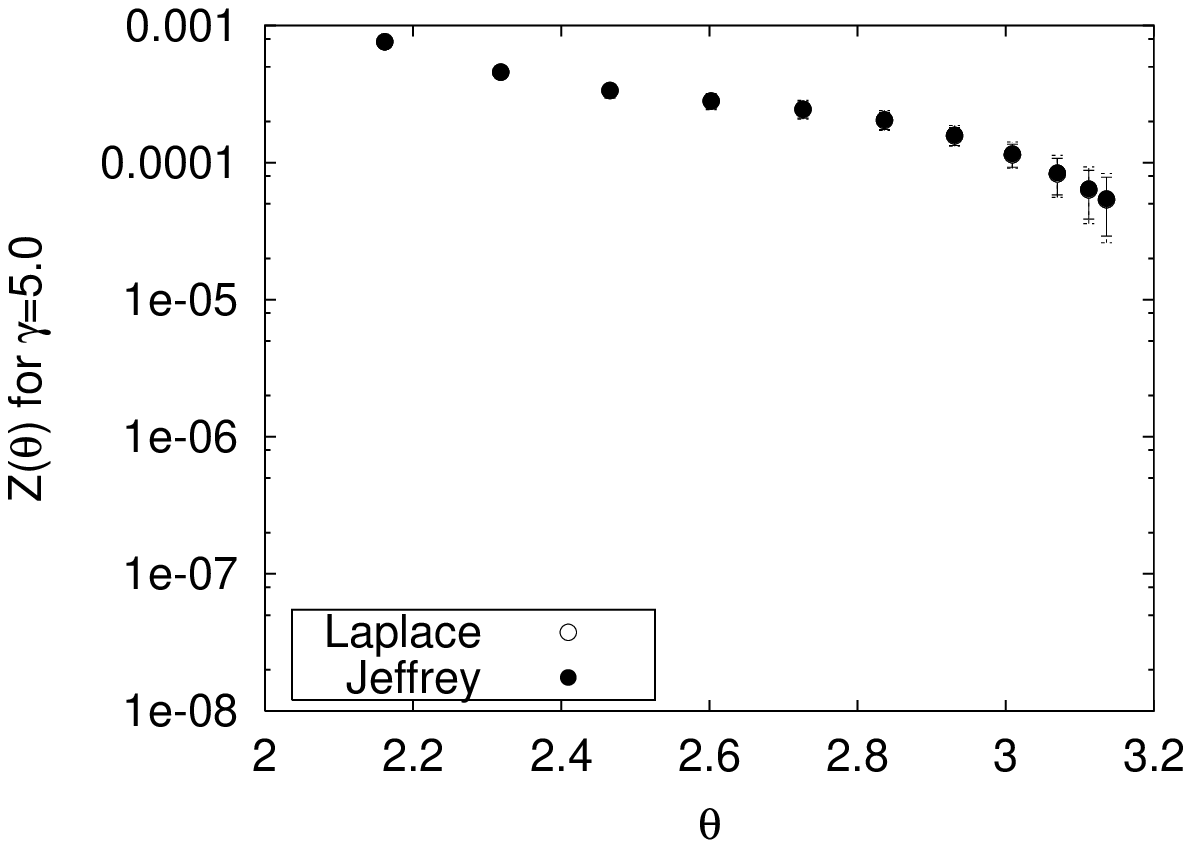}
 \end{minipage}\hfill
 \begin{minipage}{75mm}
   \includegraphics[width=75mm]{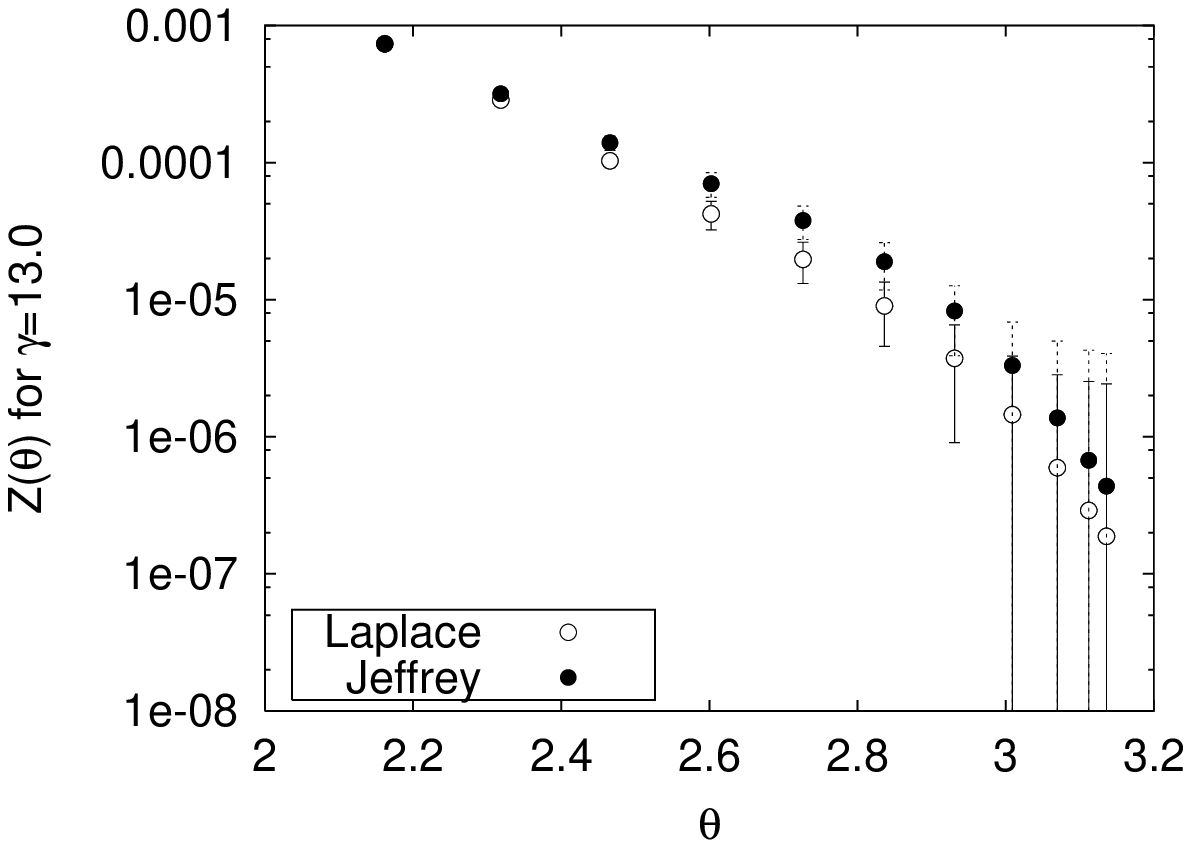}
 \end{minipage} \vspace*{-2mm}
 \caption{${\hat{\cal Z}}_{\rm Lap}(\theta)$ and ${\hat{\cal
 Z}}_{\rm Jef}(\theta)$ for $\theta\in[2.0,\pi]$~\cite{rf:ISY3}.}
 \label{fig:finalZ_LJ}
 \vspace*{-1mm}
\end{figure}

\begin{figure}[th]
 \centerline{\includegraphics[height=60mm]{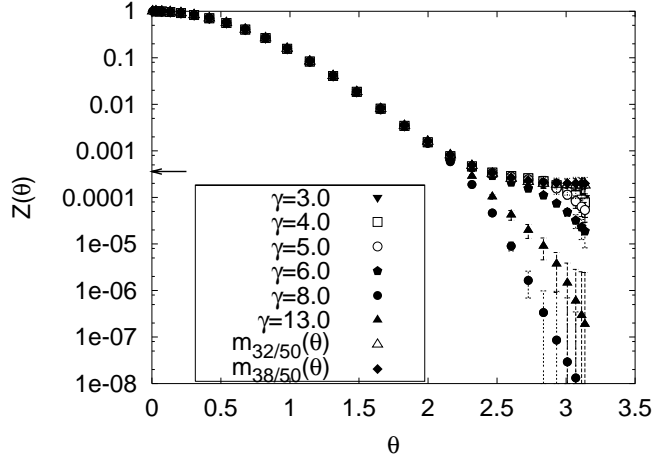}}
 \vspace{-3mm}
\caption{The most probable images ${\hat{\cal Z}}(\theta)$ for various
 $m(\theta)$~\cite{rf:ISY3}.}
\label{fig:finalZ}
\end{figure}

In this subsection, we apply the MEM to real Monte Carlo data by simulating the CP$^{3}$ model.
(For details,  see \cite{rf:ISY3}.)
For this we used a f\/ixed point action~\cite{rf:HN, rf:BISY} and various lattice volumes $L\times L$. Among these, we concentrate  on the data for $L=38$ as the non-f\/lattening case and $L=50$ as the f\/lattening case.
We systematically studied the f\/lattening phenomenon by adopting  a variety of default model $m(\theta)$  and prior probability of the parameter $\alpha$.
For the latter,  $g(\alpha)$ dependence appearing in
\begin{gather*}
 {\rm prob}(\alpha|P(Q),I,m)\equiv P(\alpha)\propto g(\alpha)
  e^{W(\alpha)+\Lambda(\alpha)}, 
\end{gather*}
is investigated.  The function $g(\alpha)$
 represents  the prior probability of $\alpha$ and is chosen according to
 prior information. In general, two types of
 $g(\alpha)$ are employed, one according to  Laplace's rule, $g_{\rm
 Lap}(\alpha)={\rm const}$, and one according to Jef\/frey's rule, $g_{\rm
 Jef}(\alpha)=1/\alpha$. The latter rule is determined by requiring that
 $P(\alpha)$  be invariant with respect to a change in scale,
 because $\alpha$ is a scale factor. The former rule  means that we have
 no knowledge about the prior information of $\alpha$.
 In general, the most probable image ${\hat{\cal Z}}(\theta)$ depends on
 $g(\alpha)$. We investigate
 the sensitivity of  ${\hat{\cal Z}}(\theta)$  to the
 choice of $g(\alpha)$ by studying a
 relative dif\/ference
\begin{gather}
 \Delta(\theta)\equiv\frac{|{\hat{\cal Z}}_{\rm Lap}(\theta)-
  {\hat{\cal Z}}_{\rm Jef}(\theta)|}
  {({\hat{\cal Z}}_{\rm Lap}(\theta)+{\hat{\cal Z}}_{\rm Jef}(\theta))/2},
  \label{eqn:Delta}
\end{gather}
 where ${\hat{\cal Z}}_{\rm Lap}(\theta)$ and ${\hat{\cal Z}}_{\rm
 Jef}(\theta)$ represent  the most probable images  according to  Laplace's rule and
 Jef\/frey's rule, respectively.

In the case without f\/lattening ($L=38$), the MEM yielded images  ${\hat{\cal
       Z}}(\theta)$ that are almost independent of $m(\theta)$ and
       $g(\alpha)$. The most probable images
       ${\hat{\cal Z}}(\theta)$ are in agreement with the result of the
       FTM within the errors.\par
 In the case with f\/lattening ($L=50$),       we found that  the
       statistical f\/luctuations  of ${\hat{\cal Z}}(\theta)$ become
       smaller as  the number of measurements increases except  near
       $\theta=\pi$.
      We also found that  ${\hat{\cal Z}}(\theta)$ with large errors
       depends strongly on $g(\alpha)$ in the region of large $\theta$,
        where the $g(\alpha)$  dependence of ${\hat{\cal Z}}(\theta)$
       was  estimated using the quantity  $\Delta(\theta)$.
       For $\theta\siml 2.3$,  ${\hat{\cal Z}}(\theta)$ agrees with the result of the FTM. For  $\theta\simr 2.3$,  ${\hat{\cal Z}}(\theta)$ behaves smoothly, while
       the FTM develops f\/lattening.
    In Fig.~\ref{fig:finalZ_LJ} , we compare     ${\hat{\cal Z}}(\theta)$ in larger values of $\theta$ for
    $g_{\rm Lap}(\alpha)$ and $g_{\rm Jef}(\alpha)$ when two dif\/ferent Gaussian default models are used ($\gamma=5.0$ and $13.0$).  It is noted that $\gamma=5.0 $ is the case in which
    the smallest values of $\Delta(\theta)$ in equation~(\ref{eqn:Delta}) are  observed  in this $\theta$ region among various default models.

       Our results are summarized in Fig.~\ref{fig:finalZ}.
       All the results obtained using the MEM behave smoothly over the entire range of
   $\theta$. Errors  are estimated from uncertainties of the images according to equation~(\ref{eqn:uncertainty}).  For  larger values of  $\theta$, ${\hat{\cal Z}}(\theta)$ depends strongly on $m(\theta)$.    Each of these  images
 could be  a candidate for the true image.  This  $m(\theta)$ dependence of ${\hat{\cal Z}}(\theta)$ may
 ref\/lect the f\/lattening phenomenon.     If we had proper knowledge about $m(\theta)$ as prior
 information, we could identify  the true image in a probabilistic sense.
  Such  knowledge  may also allow us  to
 clarify the relationship between the default model dependence and the
 systematic error, which  is  not included in the f\/igure. This will  be a  task to be pursued at the  next stage. \par
 The MEM provides a  probabilistic
 point of view in the study of theories with the sign problem. It
  may then be worthwhile to study lattice QCD with a  f\/inite density in terms
 of the MEM.

\subsection*{Acknowledgements}
 One of the authors (H.Y.) is grateful to the organizers of the O'Raifeartaigh Symposium on Non-Perturbative and Symmetry Methods in Field Theory for providing him an opportunity to present a talk.   He also thanks them  for  their  warm hospitality. The symposium reminded him well that   he   had enjoyed  intensive   discussions   with Lochlainn and other colleagues (A.~Wipf, especially)  at DIAS.

\pdfbookmark[1]{References}{ref}
\LastPageEnding


\begin{thebibliography}{99}

\footnotesize\itemsep=0pt

\bibitem{rf:Bryan} Bryan R.K., Maximum entropy analysis of oversampled data problems, {\it Eur. Biophys. J.} {\bf 18} (1990),   165--174.

\bibitem {rf:JG} Jarrell  M.,  Gubernatis J.E., Bayesian inference and the analytic continuation of imaginary-time quantum Monte Carlo data, {\it Phys.~Rep.} {\bf 269} (1996),  133--195.

\bibitem{rf:AHN} Asakawa M.,  Hatsuda T., Nakahara Y.,  Maximum entropy analysis of the spectral functions in lattice QCD,
 {\it Prog. Part. Nuclear Phys.}  {\bf  46} (2001),  459--508, \href{http://arxiv.org/abs/hep-lat/0011040}{hep-lat/0011040}.

\bibitem{rf:ISY}  Imachi M., Shinno Y.,  Yoneyama H.,  Maximum entropy method approach to $\theta$ term, {\it Progr. Theoret. Phys.}   {\bf  111} (2004),  387--411, \href{http://arxiv.org/abs/hep-lat/0309156}{hep-lat/0309156}.

\bibitem{rf:ISY2}Imachi M., Shinno Y., Yoneyama H., True or f\/ictitious f\/lattening?: MEM and the $\theta$  term, \href{http://arxiv.org/abs/hep-lat/0506032}{hep-lat/0506032}.

\bibitem{rf:ISY3} Imachi M., Shinno Y.,  Yoneyama H., Sign problem and MEM in lattice f\/ield theory with the $\theta$ term,  {\it Progr. Theoret. Phys.} {\bf 115} (2006), 931--949,
    \href{http://arxiv.org/abs/hep-lat/0602009}{hep-lat/0602009}.

\bibitem{rf:tHooft} 't Hooft  G., Topology of the gauge condition and new conf\/inement phases in non-Abelian gauge theories,  {\it Nuclear Phys.~B}  {\bf 190} (1981), 455--478.
\bibitem{rf:CR} Cardy J.L., Rabinovici E., Phase structure of Z$_{\rm p}$  models in the presence of a $\theta$ parameter,  {\it Nuclear Phys.~B} {\bf  205}  (1982),  1--16.

\bibitem{rf:CR1}
    Cardy J. L.,  Duality and the parameter in Abelian lattice models,  {\it Nuclear Phys.~B} {\bf  205} (1982),  17--26.

\bibitem{rf:Seiberg} Seiberg N.,   Topology in strong coupling, {\it Phys. Rev. Lett.} {\bf  53} (1984),  637--640.

\bibitem{rf:BRSW} Bhanot G., Rabinovici E.,  Seiberg N.,  Woit P.,
     Lattice vacua,   {\it Nuclear Phys.~B} {\bf 230} (1984),  291--298.

\bibitem{rf:Wiese}  Wiese U.-J.,  Numerical simulation of lattice $\theta$-vacua: the 2-d $U(1)$ gauge theory as a test case,  {\it Nuclear Phys.~B}  {\bf 318} (1989),  153--175.

\bibitem{rf:PS} Plefka J.C., Samuel S.,  Monte Carlo studies of two-dimensional systems with a $\theta$  term,  {\it Phys. Rev. D} {\bf  56} (1997),  44--54, \href{http://arxiv.org/abs/hep-lat/9704016}{hep-lat/9704016}.

\bibitem{rf:IKY} Imachi M.,   Kanou  S., Yoneyama H.,  Two-dimensional CP$^2$ model with $\theta$ term and topological charge distributions,  {\it Progr. Theoret. Phys.} {\bf 102} (1999),  653--670,
    \href{http://arxiv.org/abs/hep-lat/9905035}{hep-lat/9905035}.


\bibitem{rf:HN} Hasenfratz P., Niedermayer F., Perfect lattice action for asymptotically free theories,  {\it Nuclear Phys. B} {\bf 414} (1994),  785--814,\href{http://arxiv.org/abs/hep-lat/9308004}{hep-lat/9308004}.
\bibitem{rf:BISY} Burkhalter R., Imachi M.,  Shinno Y.,  Yoneyama H., CP$^{N-1}$ models with $\theta$ term and f\/ixed point action
 {\it Progr. Theoret. Phys.} {\bf 106} (2001),   613--640, \href{http://arxiv.org/abs/hep-lat/0103016}{hep-lat/0103016}.
\end{thebibliography}
\end{document}